\documentclass[12pt,a4paper]{cibb}

\makeatletter
\providecommand{\@ordinalM}[2]{#1}
\makeatother

\usepackage{subfigure,graphicx}
\usepackage{amsmath,amsfonts,latexsym,amssymb,euscript,xr}
\usepackage{booktabs}
\usepackage[nodayofweek]{datetime}
\usepackage{hyperref}
\usepackage{fmtcount}
\usepackage[english]{datenumber}
\usepackage[absolute]{textpos}
\usepackage{algorithm}
\usepackage{algorithmic}
\usepackage{color,soul}
\usepackage{comment}
\usepackage{tcolorbox}
\usepackage{xcolor}
\usepackage{chngcntr}
\usepackage{footmisc}
\definecolor{myblue}{RGB}{0, 102, 204}

\newcounter{takeaway}[section] 

\usepackage[table]{xcolor}
\usepackage{color,colortbl,tabularx}

\usepackage[english]{babel}
\usepackage[protrusion=true,expansion=true]{microtype}
\usepackage{amsmath,amsfonts,amsthm}
\usepackage{pifont}
\usepackage{fontawesome7}

\def\blue{\color{blue}}

\definecolor{LightBlue}{rgb}{0.88,0.9,0.9}

\title{\Large $\ $\\ \bf CudaMon: An R Package to Monitor NVIDIA GPUs, Showcased by Monitoring a GPU-accelerated Single-cell Analysis Workflow in R}

   \author{\blue \large Mohammad Amin Zadenoori$^1$, Riccardo Ceccaroni$^{1}$,  Gabriele sales$^{2}$, and Davide Risso$^{1}$}
\address{\blue \footnotesize $\ $\\$^1$ Department of Statistical Sciences, University of Padova, 35121 Padova, Italy \\
$^2$ Department of Biology, University of Padova, 35131, Padova, Italy
 \\

\bigskip
ORCID codes: MAZ 0000-0003-4591-153X; RC 0000-0001-5786-2830; GS 0000-0003-2078-5661; DR 0000-0001-8508-5012.
\bigskip
\newline
$^*$corresponding author: Amin.Zadenoori@Unipd.it
\bigskip
\newline
}

\abstract{\small GPU monitoring, NVIDIA GPUs, R package, performance analysis, reproducible computing, single-cell analysis. \normalsize
\\[17pt]
{\bf Abstract.} NVIDIA GPUs have recently started to be used in computational biology, yet R users lack integrated GPU monitoring tools, forcing reliance on external utilities like \texttt{nvidia-smi}. We introduce \texttt{CudaMon}, an R package providing real-time monitoring of GPU utilization, memory, temperature, and power draw via NVML, with data export and visualization utilities. Monitoring a GPU-accelerated single-cell RNA-seq pipeline (1M brain cells, RAPIDS workflow) shows compute-intensive steps (PCA, UMAP, t-SNE) exceed 90\% GPU utilization, while data management phases reveal bottlenecks. \texttt{CudaMon} facilates resource optimization, performance debugging, and reproducibility for GPU-accelerated R workflows.}
\begin{document}

\renewcommand{\thefootnote}{}
\footnotetext{\small{Article version: \datedate $\;$ h\currenttime  $\;$ CET}}

\section{Introduction}
\label{sec:SCIENTIFIC-BACKGROUND}

The increasing adoption of Graphics Processing Units (GPUs) for high-performance and parallel computing, especially in single-cell analysis\cite{dicks2026gpuacceleratedsinglecellanalysisscale,10.1145/3731599.3767378}, has made efficient resource utilization a critical concern in modern data analysis workflows. In particular, NVIDIA GPUs have started to be integrated in single-cell analysis workflows \cite{dicks2026gpuacceleratedsinglecellanalysisscale,santacatterina2025scalable}. However, users of the R programming language—one of the most widely used environments for statistical computing and bioinformatic data analysis still miss native tools to monitor GPU activity directly from their R sessions.

To address this gap, we introduce \texttt{CudaMon}, an R package designed to monitor NVIDIA GPU activity through the NVIDIA Management Library (NVML)\cite{nvml_docs}. \texttt{CudaMon} provides a simple yet powerful interface for collecting resource traces—such as GPU utilization, memory usage, temperature, and power consumption—directly from within R. 
The collected traces can be easily plotted and analyzed, enabling users to gain insights into GPU behavior without leaving their R environment.
By providing native, lightweight GPU monitoring capabilities within R, \texttt{CudaMon} empowers users to optimize resource usage, enhance reproducibility, and debug performance issues in GPU-accelerated workflows.

In this work, our contributions are twofold: (i) we design and implement an R‑based GPU‑monitoring solution built on NVIDIA’s NVML library, and (ii) we develop an end‑to‑end single‑cell analysis workflow in R that leverages Reticulate~\cite{reticulate} to interface with RAPIDS‑accelerated methods~\cite{dicks2026gpuacceleratedsinglecellanalysisscale} originally developed in Python. For clarity and reproducibility, the CudaMon R package as well as all the scripts to reproduce the results of this work are available at \cite{Anonymous2026}.

The remainder of this paper is organized as follows: Section~2 describes the CudaMon package and its usage; Section~3 presents our experimental setup and hardware specifications; Section~4 reports results from CPU–GPU runtime comparison, resource profiling, and scaling studies; and Section~5 concludes with future work directions.

\section{Using CudaMon}

As described below typical monitoring session with CudaMon follows four steps. 
First, the sampler is started with \texttt{cm\_start()}. 
Sampling occurs once per second by default, but the period and 
GPU device can be customized, e.g., 
\texttt{cm\_start(period = 0.5, device\_index = 0)}. Second, the workload is executed. Event markers can be added 
with \\\texttt{cm\_timestamp(sampler, "event\_name")}, which add 
time stamps to align code phases with GPU activity.
Third, after the workload finishes, the sampler is stopped with 
$cm\_stop(sampler)$.
Fourth, the collected CSV data is parsed using 
\texttt{cm\_parser(sampler)}. The resulting session object 
contains device metrics, process info, event timestamps, and 
file paths. Results can be visualized using \texttt{cm\_plot\_usage(session)} or, alternatively, exported as \texttt{.csv} or \texttt{dataframe}.

\begin{verbatim}
sampler <- cm_start()
cm_timestamp(sampler, "kernel_start")
# GPU computation
cm_timestamp(sampler, "kernel_end")
cm_stop(sampler)
session <- cm_parser(sampler)
cm_plot_usage(session)
\end{verbatim}

\section{Data and Methods}
\subsection{Experiments}
We designed three experiments to evaluate the performance and scalability of the GPU-accelerated workflow in Table~\ref{tab:CPU-GPU}.  First, we compare the runtime of the GPU implementation against a CPU baseline using a fine-grained micro–step analysis that isolates the execution time of each stage in the pipeline. Second, we performed a detailed profiling study of the GPU version, measuring per-step runtime, GPU utilization, CPU utilization, and memory usage to characterize how computational load and resource consumption evolve throughout the workflow. Third, we conduct a subsampling experiment in which the input dataset size increases from 100k to 1M in increments of 100k, allowing us to quantify how total runtime and GPU memory usage grow with data size and to assess the scalability of the full pipeline.

\subsection{GPU and CPU Monitoring}
GPU utilization is monitored using CudaMon, which is called from within an R script, while CPU usage is tracked via the \texttt{top -p} command in bash. For a given instance running \texttt{Rscript scWorkflow.R \&}, the process ID (PID) is first captured. Then, the monitoring command \texttt{top -p PID -b -d 1 > continuous\_monitoring.txt} is executed. This runs \texttt{top} in batch mode (\texttt{-b}), targets the specified process (\texttt{-p PID}), and sets a 1-second update interval (\texttt{-d 1}). The output is redirected to a text file named \texttt{continuous\_monitoring.txt}, thereby saving resource metrics for the process every second. Finally, a dedicated parser extracts two key metrics from the log file: the CPU utilization percentage and the memory usage of the monitored R process throughout its execution.
\subsection{Data}
As a case study, we used a single-cell RNA-seq data of 1.3 million E18 mouse brain cells from 10X Genomics. The data consist of 1.3 million cells and 27,998 genes~\cite{Zheng2017}. In this study, after loading the entire dataset, we selected the first 1 million cells, as done in the original Rapids-singlecell tutorial.

\subsection{Hardware Specification}
The hardware system used to run the pipeline in Table~\ref{tab:CPU-GPU} consists of two distinct node types. The CPU version of the pipeline, which is based on \texttt{Scanpy} \cite{wolf2018scanpy}, was executed on a CPU node featuring 96 AMD EPYC 9654 cores (384 logical cores) and 768~GB of RAM. The GPU version of the pipeline was executed on a GPU node equipped with 2 AMD EPYC 9135 processors, 1.5~TB of CPU RAM, and an NVIDIA H200 NVL GPU with 141~GB of HBM3 memory. 

\section{Results}

\subsection{Experiment 1: CPU--GPU Runtime Comparison}

Table~\ref{tab:CPU-GPU} compares runtime and peak memory usage of each step of a typical single-cell analysis pipeline between GPU-accelerated and CPU-only implementations. The full GPU pipeline completed in 152 seconds, while the CPU pipeline required 4,659 seconds (approximately 80 minutes), representing a 32× speedup. The most dramatic improvements were observed in clustering (1938.78 CPU vs. 8s GPU). These results demonstrate that GPU acceleration reduces total analysis time from nearly 1.5 hours to under 3 minutes, enabling rapid iterative exploration of large-scale single-cell datasets.  During the CPU-based pipeline, CPU memory usage peaked at 134~GB, whereas in the GPU-based pipeline, GPU memory usage peaked at 101~GB and CPU memory usage peaked at only 64~GB.  Furthermore, the higher GPU memory usage compared to CPU memory (101~GB vs. 64~GB) in the GPU pipeline indicates that Rapids offloads the majority of computation to the GPU, leveraging its memory bandwidth and parallelism more effectively than the CPU. In the only CPU-based pipeline, the pipeline was allowed to use 200 cores, but based on the observation in \texttt{top}, it sometimes used just one, and the maximum cores used was 10.

\begin{table}[!ht]
\centering
\scriptsize
\begin{tabular}{lccc|cc}
\hline
 & \multicolumn{3}{c|}{\textbf{GPU Pipeline}} & \multicolumn{2}{c}{\textbf{CPU Pipeline}} \\
\cline{2-6}
\textbf{Phase} & \textbf{Runtime (s)} & \textbf{GPU Memory (GB)} & \textbf{CPU Memory (GB)} & \textbf{Runtime (s)} & \textbf{CPU Memory (GB)} \\
\hline
1. Loading                     & 62 & 60& 62 & 74.91 & 70 \\
2. Quality control             & 4  & 65& 62 & 86.60 & 100 \\
3. Normalization + HVG selection & 7  & 80& 64 & 46.18 & 100 \\
4. Regression / Scaling        & 1  & 80 & 64 & 40  & 100 \\
5. PCA (dimension reduction)   & 2  & 82& 64 & 129.68 & 115 \\
6. kNN graph                   & 10 & 82& 64 & 159.87 & 126 \\
7. Umap + t-sne  & 20 & 101 & 64 & 1557.68 & 126 \\
8. Louvain + Leiden clustering                & 8  & 84 & 64 & 1938.78 & 134 \\
9. Differential expression     & 20 & 101& 64 & 384.76 & 134 \\
10. Diffusion maps (trajectory) & 18 & 101 & 64 & 81.27 & 134 \\
\hline
\multicolumn{1}{r}{\textbf{Total}} & \textbf{152} & & & \textbf{4659.48 } & \\
\hline
\end{tabular}
\caption{Comparison of per-step runtime and \textbf{peak} memory usage between GPU-accelerated and CPU-only implementations of the single-cell analysis pipeline.}
\label{tab:CPU-GPU}
\end{table}
\begin{figure}[!ht] 
    \centering
    \includegraphics[width=1\textwidth]{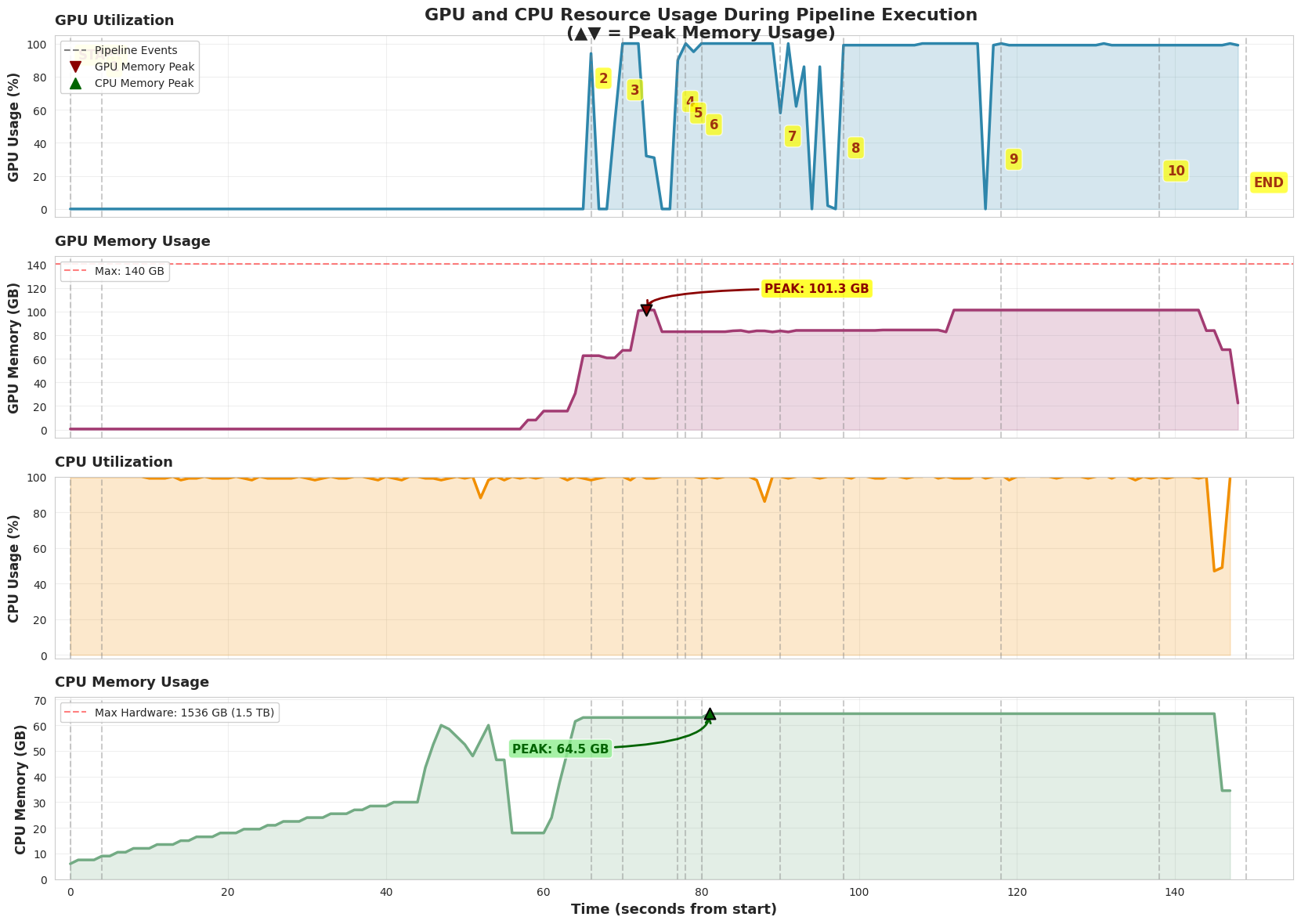}
    \caption{Stepwise resource monitoring of the GPU-accelerated workflow, where each interval between consecutive timestamps (e.g., 4--66~s as \textsc{Load}, 78--80~s as \textsc{Pca}, 90--92~s as \textsc{Umap}) is assigned to the first timestamp to delineate computational steps. The name of each steps can be mapped from the numeric step identifiers or start timestamps listed in Table~\ref{tab:CPU-GPU}.}
    \label{fig:scWorkflowMonitor}
\end{figure}

\subsection{Experiment 2: Profiling of GPU Runtime and Resource Utilization}

The computational pipeline exhibits two distinct phases: an initial CPU-bound preprocessing window ($t < 60$\,s) and a subsequent GPU-intensive execution phase ($t > 60$\,s) depicted in Figure~\ref{fig:scWorkflowMonitor}. During preprocessing, CPU utilization ($U_{CPU}$) remains saturated at $\approx 100\%$ with a step-wise increase in memory, while the GPU remains idle. The transition at $t = 60$\,s triggers a rapid escalation in GPU memory, reaching a peak ($M_{GPU,peak}$) of $101.3$\,GB at $t \approx 75$\,s, representing $72\%$ of the $140$\,GB hardware limit. Concurrently, CPU memory peaks at $64.5$\,GB, a marginal fraction of the available $1.5$\,TB capacity. In the terminal steady-state ($t > 100$\,s), $U_{GPU}$ achieves near-total compute saturation ($\approx 100\%$) while the CPU maintains maximum utilization for data orchestration until synchronized resource release at $t \approx 152$\,s. 
Based on what is shown in Figure~\ref{fig:scWorkflowMonitor}, the PCA, regression-out, KNN, differential expression, and trajectory steps are the most computationally intensive on the GPU. This outcome is expected, since these are not only the most parallelizable operations but also the ones where the most dramatic improvements, compared to CPU execution, occurred.
\begin{figure}[!ht] 
    \centering
    \includegraphics[width=1\textwidth]{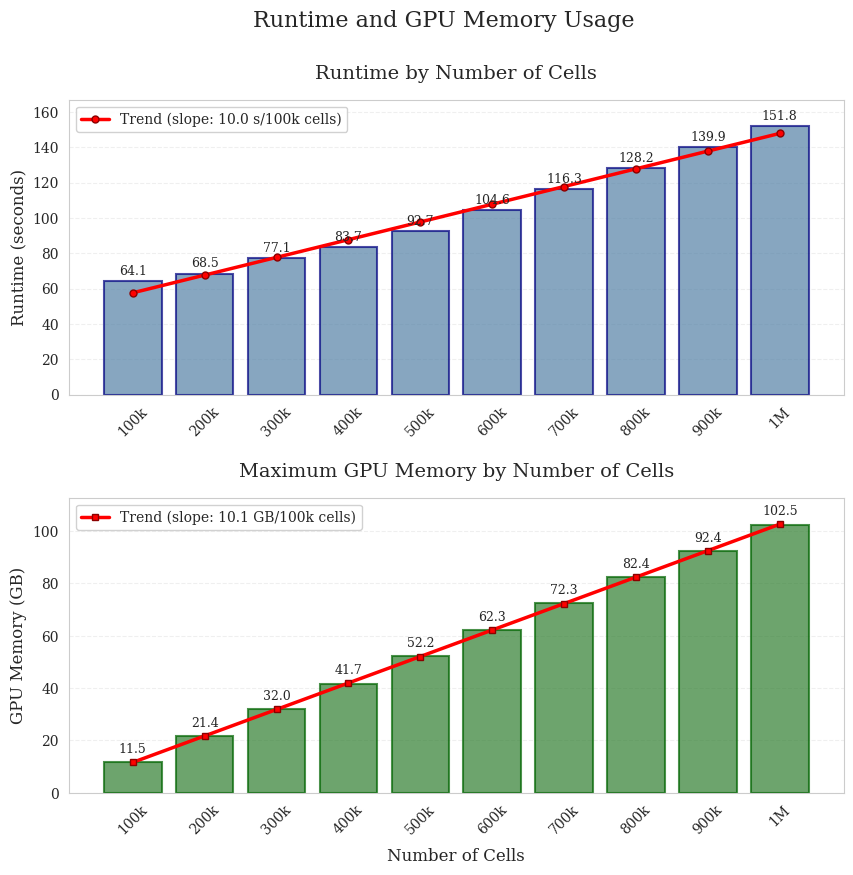}
    \caption{ Impact of cell count on computational cost. (a) Runtime in seconds showing linear growth with increasing cell count. (b) Maximum GPU memory in GB demonstrating analogous linear scaling. Trend lines (red) confirm consistent performance degradation patterns.}
    \label{fig:scaling}
\end{figure}
\subsection{Experiment 3: Scaling Study from 100k to 1M Cells}

We evaluated computational scalability by varying the number of cells from \(1 \times 10^5\) to \(1 \times 10^6\) in ten equally spaced increments (100k, 200k, …, 1M cells). For each configuration, we measured total runtime (seconds) and peak GPU memory usage (GB) of the GPU version of the pipeline in Table~\ref{tab:CPU-GPU}. As shown in Figure~\ref{fig:scaling}, both metrics increase approximately linearly with cell count. Runtime grows from 64.1~s at 100k cells to 151.8~s at 1M cells (2.4\(\times\) increase), with a linear trend of \(\sim\)9.7~s per additional 100k cells. GPU memory rises from 11.5~GB to 102.5~GB (8.9\(\times\) increase), with a slope of \(\sim\)10.1~GB per 100k cells. This scaling study reveals that while runtime remains modest at 1M cells (\(\approx\)2.5~minutes), GPU memory becomes the primary bottleneck beyond 1.2–1.5M cells. The linear trends enable accurate extrapolation to larger cell counts and provide unit costs (9.7~s and 10.1~GB per 100k cells) for resource planning.

\section{Conclusions}

In this work, we introduced CudaMon, an R package that provides native, lightweight GPU monitoring capabilities by interfacing directly with NVIDIA's NVML library. CudaMon enables R users to collect real-time resource traces, including GPU utilization, memory usage, temperature, and power draw, without leaving the R environment, addressing a critical gap in the R ecosystem for GPU-accelerated computing. The package's four-step workflow (start, timestamp, stop, parser) and built-in visualization utilities make performance monitoring accessible to computational biologists and statisticians alike.

While in the current work we use \texttt{top} to monitor CPU resources, in the future we plan to integrate CudaMon with the Rcollectl Bioconductor package \cite{Rcollectl}. This will provide R users with a set of monitoring tools for joint CPU-GPU profiling from within R.

\label{sec:CONFLICT-OF-INTERESTS}

\section*{Acknowledgments (optional)}
\label{sec:ACKNOWLEDGMENTS}
The authors would like to thank Vince J. Carey for his helpful feedback on a preliminary version of the package.

\section*{Funding (optional)}
\label{sec:FUNDING}
This work was supported in part by project EOSS6-0000000644 from the Chan Zuckerberg Initiative and by the European Research Council (ERC) Grant CoG 101171662.
\section*{Availability of data and software code (optional and strongly suggested)}
\label{sec:AVAILABILITY}
For clarity and reproducibility, the CudaMon R package as well as all the scripts to reproduce the results of this work are available at \cite{Anonymous2026}.

\footnotesize
\bibliographystyle{unsrt}
\bibliography{bibliography_CIBB_file.bib} 
\normalsize

\end{document}